\documentclass[12ppt]{emulateapj}
\usepackage{natbib}
\usepackage{color}
\tightenlines

\def \bea {\begin{eqnarray}}
\def \ena {\end{eqnarray}}

\begin{document}
\title{Acceleration of small astrophysical grains due to charge fluctuations}
\author{A. V. Ivlev$^1$, A. Lazarian$^2$, V. N. Tsytovich$^3$,  U. de Angelis$^4$, Thiem Hoang$^2$, G. E.~Morfill$^1$}
\email[E-mail:~]{ivlev@mpe.mpg.de}

\affiliation{$^1$Max-Planck-Institut f\"{u}r extraterrestrische Physik, 85741 Garching, Germany\\
$^2$Department of Astronomy, University of Wisconsin, Madison, WI 53706, USA\\
$^3$General Physics Institute, Russian Academy of Sciences, 117942 Moscow, Russia\\
$^4$Department of Physical Sciences, University of Naples and INFN, sezione di Napoli, Italy}

\date{\today}

\begin{abstract}
We discuss a novel mechanism of dust acceleration which may dominate for particles smaller than $\sim0.1~\mu$m. The
acceleration is caused by their direct electrostatic interactions arising from fluctuations of grain charges. The energy
source for the acceleration are the irreversible plasma processes occurring on the grain surfaces. We show that this
mechanism of charge-fluctuation-induced acceleration likely affects the rate of grain coagulation and shattering of the population
of small grains.

\end{abstract}


\maketitle

\section{Introduction}
Dust is an important constituent of interstellar medium (ISM), molecular clouds and accretion disks (see Whittet 2003,
Draine 2009). It gets involved in many key processes, for instance, it controls heating and cooling of the ISM (see Draine
2003, Tieliens 2005), reveals magnetic fields through grain alignment (see Lazarian 2007 for a review) and interferes with
the attempts to measure properties of CMB radiation (see Lazarian \& Finkbeiner 2003, Fraisse et al. 2009).

Small, i.e., less than $\sim10^{-5}$~cm grains, are an important component of the interstellar dust population, with a
notable fraction of very small grains -- Polycyclic Aromatic Hydrocarborn (PAH) particles, which are essentially large
molecules\footnote{The information about these grains is being obtained through the transient heating of grains by UV
photons. Indeed, after absorbing a photon a grain gets heated to rather high temperature, inducing emission of infrared
photons (see Allamandola et al. 1989). The PAHs were invoked by Draine \& Lazarian (1998) to explain the anomalous emission
of observed in the range of 10-100 GHz (see also Hoang, Draine \& Lazarian 2010).} (see Leger \& Puget 1984). In what
follows, due to the reasons that are explained in \S 5, we do not directly address PAH particles, but our approach to the
small grain acceleration may be extended to this important population of grains.

Most properties of grains, including light extinction, electron photoemission, and chemical activity depend not only on
grain chemical composition, but also on their sizes. In astrophysical media, these sizes are affected by grain-grain
collisions. The minimal velocities of grains are determined by their Brownian motion corresponding to the temperature of the
ambient gas. Large-scale hydrodynamic motions associated with turbulence can make grains move faster (see Draine 1985).

Since most astrophysical media are magnetized and grains are charged the hydrodynamic treatment of acceleration is
frequently not adequate. A proper treatment of the grain acceleration through the interaction of charged grains with
magnetohydrodynamic (MHD) turbulence has been developed recently (Lazarian \& Yan 2002, Yan \& Lazarian 2003, Yan, Lazarian
\& Draine 2004, Yan 2009). This treatment makes extensive use of the advances of compressible MHD turbulence (see Cho \&
Lazarian 2002, 2003) and provides the mathematical formalism of the second-order Fermi acceleration of charged grains
interacting with MHD turbulence.

However, the acceleration mechanisms based on the MHD interaction of turbulence and charged grains exhibit acceleration
rates that decrease as dust particles get smaller. This decrease arises from the fact that the Larmor radius of charged
grains becomes smaller with the decrease of grain mass and, correspondingly, grains have to interact with smaller, i.e.,
less powerful, turbulent fluctuations. In addition, compressible fluctuations, i.e., fast modes, which were identified in
Yan \& Lazarian (2003) with the most efficient acceleration, get suppressed at the small scales due to plasma damping, while
the Alfvenic mode gets inefficient for acceleration at small scale due to anisotropy (see Yan \& Lazarian 2003 for more
discussion). The acceleration of grains with sizes less than $\sim10^{-5}$~cm becomes rather inefficient for most media
discussed in Yan \& Lazarian (2003), Yan, Lazarian \& Draine (2004, henceforth YLD04).

Are there other mechanisms of dust acceleration which dominate for grains smaller than $10^{-5}$~cm? One can expect
that dust-plasma interactions may be important for such grains. This paper presents a novel promising acceleration mechanism
based on grain-grain Coulomb collisions in the presence of grain charge fluctuations. This mechanism utilizes intrinsic
non-equilibrium nature of the dusty astrophysical plasmas. In addition to acting on the small grains, the mechanism may
drive the acceleration of larger grains whenever the MHD acceleration mechanism is suppressed (e.g., when MHD turbulence is
damped).

In what follows, we introduce major timescales characterizing dynamics of charged interstellar grains in \S2, present the
mechanism of small grain acceleration associated with charge fluctuations during their Coulomb collisions in \S3, calculate
the acceleration for grains of a given size distribution in \S4, analyze implications of the effect for interstellar phases
in \S5, discuss the importance of our results in \S6, and summarize them in \S7.

\section{Dust interactions in interstellar gas}

\subsection{Dust damping time}

Interactions of dust with the ambient gas present the primary mechanism of dissipating streaming motions of grains. The
damping rate of translational motion arising from the interaction with neutral gas is essentially the inverse time for
collisions with the mass of the gas equal that of a grain (Purcell 1969),
\begin{equation}
\tau_{dn}^{-1}= 2\sqrt{\frac2{\pi}}\frac{n_n}{a\rho_d}(m_nk_{\rm B}T_n)^{1/2},
\label{tau_fr}
\end{equation}
where $m_n$, $n_n$, and $T_n$ are the mass, volume density, and temperature of neutrals, $\rho_d$ is the mass density of
dust grains and $a$ is their radius.

When the ionization is sufficiently high, the interaction of charged grains with a plasma becomes important (Draine \&
Salpeter 1979). The ion-grain cross section due to long-range Coulomb force is larger than the atom-grain cross section. As
a result, the rate of translational motion damping gets modified (Draine \& Salpeter 1979). For subsonic motions the
effective damping rate is renormalized, $\tau_{\rm damp}^{-1}=\alpha\tau_{dn}^{-1}$, with the following renormalizing
factor:
\begin{eqnarray}
\alpha=1+\frac{n_{\rm H}}{2n_n}\sum_{i}x_i\left(\frac{e^{2}}{ak_{\rm B}T_i}\right)^{2}\left(\frac{m_{i}}{m_n}\right)^{1/2}
\sum_{Z}Z^2f(Z)\nonumber\\
\times\ln\frac{3}{2\sqrt{\pi}}\frac{(k_{\rm B}T_i)^{3/2}}{|Z|e^3(xn_{\rm H})^{1/2}}.\hspace{.5cm}\label{alpha1}
\end{eqnarray}
Here $x_{i}$ is the abundance of ion $i$ (relative to hydrogen) with mass $m_{i}$ and temperature $T_i$, $x=\sum_i x_i$, and
$f(Z)$ is the grain charge distribution function (see Hoang, Draine \& Lazarian 2010). When the grain velocity $v_d$
relative to the gas becomes supersonic the dust interactions with the plasma is diminished, and the damping rate in this
case is renormalized due to the gas-dynamic correction (Purcell 1969),
\begin{equation}
\alpha=\left(1+\frac{9\pi}{128}\frac{v_d^2}{C_{\rm s}^2}\right)^{1/2},\label{alpha2}
\end{equation}
where $C_{\rm s}=\sqrt{k_{\rm B}T/m_n}$ is the sound speed.

The Larmor rate for a grain is $\tau_{\rm L}^{-1}=|Z_0|eB/m_{d}c$, where $Z_0(a)$ is the mean grain charge and $B$ is the
magnetic field strength. If $\tau_{\rm L}$ is larger than $\tau_{\rm damp}$ the effect of magnetic field on dust dynamics is
negligible.

\subsection{Grain Coulomb interactions}

\begin{figure}
\includegraphics[width=0.5\textwidth]{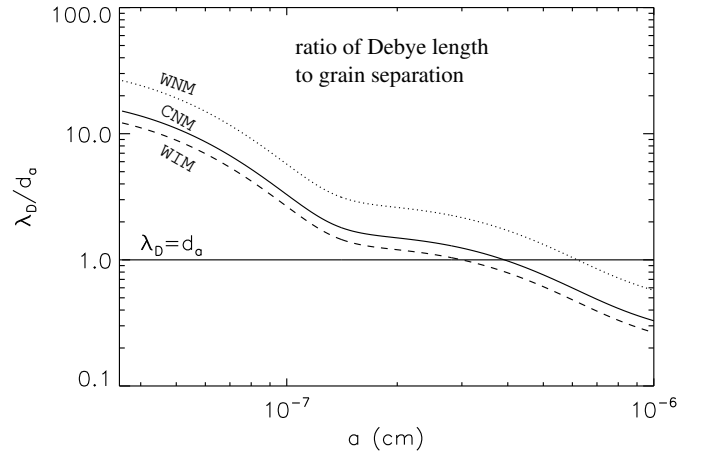}
\caption{The ratio of the Debye screening length, $\lambda_{\rm D}$, to the distance $d_a$ between grains larger than a given size $a$.}
\label{f1}
\end{figure}

Grains can interact with each other through Coulomb forces if the Debye screening length is larger than the distance between
grains. In Figure~\ref{f1} we show the ratio of the Debye radius $\lambda_{\rm D}$ to the separation $d_a$ between the grain
of size $a$ and grains of a larger size, calculated for three different idealized interstellar phases: the cold neutral
medium (CNM), warm neutral medium (WNM), warm ionized medium (WIM). Table \ref{tab1} presents physical parameters for such
environments [including also reflection nebula (RN) and photodissociation region (PDR)], where $n_{\rm H}$ is the hydrogen
density, $T$ and $T_{d{\rm s}}$ are the gas temperature and dust material temperature, respectively, $n({\rm H}^{+}), n({\rm
M}^{+})$ and $n({\rm H}_{2})$ are ion hydrogen density, ion metal density and molecular hydrogen density, respectively. To
represent grain charge we use the charging data from Hoang et al. (2010).

It is clear that for grains smaller than $\sim 3\times 10^{-7}$~cm the grains do interact electrostatically with their
larger neighbors. As a result of charge fluctuations, the dust particles experience forces which can accelerate them.

We choose to depict distances to the grains of the size larger than the given grain size, because such interactions will be
most important in terms of momentum transfer. Naturally, larger grains will be still interacting with smaller ones and
therefore there should be no sharp cut-off for the grains to stop being accelerated by the charge-fluctuation mechanism.

\begin{table}
\caption{Idealized Environments for the Interstellar Matter}\label{tab1}
\begin{tabular}{llllll} \hline\hline\\
\multicolumn{1}{c}{Parameter} & \multicolumn{1}{c}{CNM}& {WNM} &WIM &RN &PDR\\[1mm]
\hline\\
$n_{\rm H}({\rm cm}^{-3})$ &30 &0.4 &0.1 &$10^{3}$ &$10^{5}$ \\[1mm]
$T$(K)& 100& 6000 &8000 &100 &1000\\[1mm]
$T_{d{\rm m}}$(K)& 20& 20 &20 &40 &80\\[1mm]
$x_{\rm H}=n({\rm H}^{+})/n_{\rm H}$ &0.0012 &0.1 &0.99 &0.001 &0.0001\\[1mm]
$x_{\rm M}=n({\rm M}^{+})/n_{\rm H}$ &0.0003 &0.0003 &0.001 &0.0002 &0.0002\\[1mm]
$y=2n({\rm H}_{2})/n_{\rm H}$&0. &0. &0. &0.01 &0.01\\[1mm]
\\[1mm]\hline\hline\\
\end{tabular}
\end{table}

\section{Mechanism of dust acceleration}

\subsection{Physics of the process}

Astrophysical grains are known to be charged (see Draine \& Sutin 1987). The grain charges are not constant -- they
fluctuate around an equilibrium value $Q_0=eZ_0$. In a plasma without external radiation $Q_0$ is negative, because
electrons are much faster than ions (in the presence of sufficiently intense external UV radiation it becomes positive, see
Goree 1994). In general, one needs to deal with a distribution of charges (see Weingartner \& Draine 2001, Hoang et al.
2010).

It is possible to demonstrate that the mutual Coulomb collisions between dust particles with fluctuating charges do not
conserve the kinetic energy of the particles. As the energy is being taken from the fluctuations the kinetic energy of the
particles increase in a stochastic way (see de Angelis et al. 2005, Ivlev et al. 2005, henceforth I05) -- a process similar
to second-order Fermi acceleration, where the role of randomly moving magnetic mirrors is played by the randomly fluctuating
electric forces exerted by other particles.

The physical mechanism behind such stochastic heating is easy to understand. Each particle creates in its vicinity a
fluctuating electric field, so that the superposition of these fluctuations creates a stochastic background which can be
stronger than the fluctuations induced by the plasma itself. An individual dust particle in such a random environment
experiences fluctuating electric force and hence performs a random walk in the momentum space. This naturally results in the
stochastic acceleration of dust particles.

Below we elaborate on this mechanism. We first consider binary interactions (collisions) between particles with fluctuating
charges, and demonstrate that such interactions result in the effective particle acceleration. Next, we employ the kinetic
approach and show that in large particle ensembles the stochastic acceleration in binary collisions can result in the
exponential growth of the particle kinetic energy (temperature). We stress that we consider collisions between grains only
for illustrative purposes, as it is one way to show that energy of grains do change as a result of charge fluctuations. At
the same time, the essential element of the acceleration process is the fluctuation of electric force acting on a particle,
which occurs even if grains are not directly colliding.

\subsection{Stochastic acceleration in binary collisions}

We consider dilute dust clouds and hence only focus on the binary interactions between grains. Binary collisions can be
conveniently studied in terms of the center-of-mass and relative coordinates. Below we consider grains of the same mass,
$m_{d1}=m_{d2}(=m_d)$, although all results can be straightforward generalized for arbitrary mass ratio (Ivlev et al. 2004):
e.g., for $m_{d1}/m_{d2}\ll1$ one should substitute the mass of light particle instead of the reduced mass ($\frac12m_d$).
For a pair of particles with momenta ${\bf p}_1$ and ${\bf p}_2$, the center-of-mass and relative momenta are ${\bf p}_{\rm
c}=\frac12({\bf p}_1+{\bf p}_2)$ and ${\bf p}_{\rm r}={\bf p}_1-{\bf p}_2$, respectively. The kinetic energy of the pair can
be expressed via $p_{\rm c}$ and $p_{\rm r}$ as follows: $p_1^2/2m_d+p_2^2/2m_d=p_{\rm c}^2/m_d+p_{\rm
r}^2/4m_d\equiv\varepsilon_{\rm c}+\varepsilon_{\rm r}$. The center-of-mass momentum and hence the energy $\varepsilon_{\rm
c}$ are conserved during the collision (here we neglect momentum exchange due to collisions with electrons and ions),
whereas the relative momentum is changed,
\begin{displaymath}
{\bf p}_{\rm c}'={\bf p}_{\rm c};\qquad {\bf p}_{\rm r}'={\bf p}_{\rm r}+{\bf q}.
\end{displaymath}
For constant charges $\varepsilon_{\rm r}$ is obviously conserved during the collision, but when charges vary
$\varepsilon_{\rm r}$ varies as well. Thus, in the presence of charge fluctuations the exchange of the relative momentum can
be divided into the elastic and inelastic parts, ${\bf q}={\bf q}_0+\delta{\bf q}$. The elastic part keeps the magnitude of
the relative momentum constant, $|{\bf p}_{\rm r}+{\bf q}_0|=|{\bf p}_{\rm r}|$, so that the energy variation after the
collision is $\delta\varepsilon_{\rm r}=({\bf p}_{\rm r}+{\bf q}_0)\cdot\delta {\bf q}/2m_d+(\delta {\bf q})^2/4m_d$. The
inelastic momentum exchange $\delta {\bf q}$ is generally a function of ${\bf p}_{\rm c}$ and ${\bf p}_{\rm r}$ and is
determined by the stochastic properties of charge fluctuations.

In order to calculate the energy variation, we suppose that for different particles the charge fluctuations are
uncorrelated, so that for a pair of particles with charges $Q_1(t)=Q_0+\delta Q_1(t)$ and $Q_2(t)=Q_0+\delta Q_2(t)$ we have
$\langle\delta Q_1(t+\tau)\delta Q_2(t)\rangle=0$. Assuming that $|\delta Q/Q_0|\ll1$, the inelastic momentum exchange due
to the collision can be calculated as $\delta{\bf q}\approx\int[\delta Q_1(t)+\delta Q_2(t)]\nabla\varphi_0dt$, where
$\varphi_0(r)$ is the potential of the equilibrium charge $Q_0$ and the integration is performed along the (equilibrium)
collision trajectory ${\bf r}(t)$. The resulting energy variation is obtained by averaging over the fluctuations, i.e.,
\begin{equation}\label{4}
\delta\varepsilon_{\rm r}\equiv\frac{\langle(\delta {\bf q})^2\rangle}{4m_d}=\frac1{2m_d}\left\langle\int\delta
Q\nabla\varphi_0dt\int\delta Q'\nabla\varphi_0'dt'\right\rangle.
\end{equation}
To calculate $\delta\varepsilon_{\rm r}$ we also assume $|\delta Q/Q_0|\ll1$ and therefore can use the following charge
auto-correlation function (Matsoukas \& Russell 1997):
\begin{equation}\label{24}
\langle\delta Q(t+\tau)\delta Q(t)\rangle=\sigma_Q^2{\rm e}^{-\nu_{\rm ch}|\tau|}.
\end{equation}
Here $\sigma_Q^2$ is the charge dispersion and $\nu_{\rm ch}$ is the ``charging frequency'' (which plays the role of the
scale parameter in the charge fluctuations spectrum). The assumption about small charge fluctuations simplifies
significantly the derivation of Eq. (\ref{4}), and also allows us to treat the charge as a continuous variable and hence
employ Eq. (\ref{24}).

There are several mechanisms that can result in the fluctuations (Yan et al. 2004, Khrapak et al. 1999). In this paper, for
the sake of convenience we focus on the {\it most elementary} charging process which is due to discreteness of the
elementary charge. Usually, this is a Gaussian process with $\sigma_Q^2=\frac{1+z}{z(2+z)}|eQ_0|$ and $\nu_{\rm
ch}=\frac1{\sqrt{2\pi}}(1+z)a\omega_{{\rm p}i}^2/v_{T_i}$ (Matsoukas \& Russell 1997), where $\omega_{{\rm p}i}=\sqrt{4\pi
e^2n_i/m_i}$ is the ion plasma frequency, $v_{T_i}=\sqrt{k_{\rm B}T_i/m_i}$ is the ion thermal velocity, and
$z=e|Q_0|/ak_{\rm B}T_i$ is the dimensionless charge of dust particle of the radius $a$ (for different gases, $z=1-3$)
(Tsytovich 1997, Fortov et al. 2005). Note that other charging processes can also play an important role (in particular, the
photoelectric emission) and require separate careful consideration.

One can identify two limiting collisional regimes associated with the charge fluctuations. In the limit of {\it rapid
fluctuations} the charges vary many times in each act of binary collisions between dust particles. This regime corresponds
to $\nu_{\rm ch}\tau_{\rm int}\gg1$, where $\tau_{\rm int}(v_d)$ is the characteristic timescale of the dust-dust
interaction during the collision. By employing Eq. (\ref{24}), we readily obtain from Eq. (\ref{4}) the following limiting
expression for the energy variation:
\begin{equation}\label{25}
\delta\varepsilon_{\rm r}\approx\frac{\sigma_Q^2}{m_d\nu_{\rm ch}}\int\left|\frac{d\varphi_0}{dr}\right|^2dt+O(\nu_{\rm ch}^{-2}).
\end{equation}
Equation (\ref{25}) shows that in this limit each collision between dust particles results in the increase of their kinetic
energy (I05).

The interaction timescale $\tau_{\rm int}$ is determined by the range of the fluctuating field. In contrast to the average
field, which is screened at the plasma Debye length $\lambda_{{\rm D}}$, the screening of the fluctuating field occurs at
the so-called ``plasma flux'' scale $\lambda_{\rm flux}$ (see, e.g., Tsytovich et al 2008). The physical meaning of
$\lambda_{\rm flux}$ is rather simple: While the screening of the average field is characterized by the timescale
$\sim\omega_{{\rm p}i}^{-1}$ and occurs at the length scale $\sim v_{T_i}/\omega_{{\rm p}i}=\lambda_{{\rm D}}$, the relevant
timescale for charge fluctuations is $\sim\nu_{\rm ch}^{-1}$ and hence they are screened at the length scale
$\sim\lambda_{\rm flux}= v_{T_i}/\nu_{\rm ch}$. Thus, the dust interaction timescale is $\tau_{\rm int}\sim\lambda_{\rm
flux}/v_d$ and therefore the condition of rapid fluctuations is reduced to $v_{T_i}/v_d\gg1$, i.e., it is generally
satisfied for the subsonic dust.

In the opposite limit of {\it slow fluctuations} the characteristic time of the charge variations is much longer than the
collision time. This regime is completely analogous to ``regular'' Fermi acceleration -- with equal probability, each
collision results in energy gain or loss, but the collisions resulting in the energy gain are more frequent. In this case,
after averaging over many collisions, we also obtain net energy growth. Analysis of this regime will be presented elsewhere.

\subsection{Evolution of dust kinetic temperature}

In order to understand the effect of the energy variation in binary dust-dust collisions [given by Eq. (\ref{25})] on the
mean kinetic energy of the whole dust ensemble one should employ the kinetic approach. The kinetics is described in terms of
the velocity distribution function $f_d({\bf p},t)$. There are two principal contributions to the dust kinetics -- one is
due to the mutual dust collisions and another because of dust interactions with the ambient gas. For simplicity, we first
consider the idealized situation for the second contribution and only take into account the interactions with neutrals of
temperature $T_n$ (we generalize this idealized setup for more realistic environments in \S 5).

The resulting kinetic equation has the following form:
\begin{equation}\label{01}
\frac{df_{d}}{dt}={\rm St}_{dd}f_{d}+{\rm St}_{dn}f_d,
\end{equation}
where St$_{dd}$ and St$_{dn}$ denote the collision operators (integrals) describing the dust-dust and dust-gas interactions,
respectively. Although the analysis of Eq. (\ref{01}) can be performed for arbitrary $f_d({\bf p})$, for the sake of
convenience we assume that dust particles have a Maxwellian velocity distribution $f_{\rm M}({\bf p})$. Then the equation
for evolution of mean kinetic energy (kinetic temperature) $k_{\rm B}T_d(t)=\frac13\int(p^2/m_d)f_{\rm M}d{\bf p}$ is
obtained by taking the second moment of Eq. (\ref{01}),
\begin{equation}\label{4a}
\dot T_d=\frac13\int\frac{p^2}{m_d}\left({\rm St}_{dd}f_{\rm M}+{\rm St}_{dn}f_{\rm M}\right)d{\bf p}.
\end{equation}

The integrals in Eq. (\ref{4a}) are calculated separately in Appendices \ref{app1} and \ref{app2}: Equation (\ref{int1})
yields the second integral (contribution of dust-neutral interactions), whereas the first integral describing the dist-dust
collisions is given by Eq. (\ref{8}). The latter is determined by the kinetic coefficient ${\cal A}_{dd}$, which is
calculated in Appendix \ref{app3}, Eq. (\ref{27}). By substituting this in Eq. (\ref{8}) (and taking into account that the
coefficient ${\cal B}_{dd}$ can be neglected, see Appendix \ref{app2}), we derive the following equation for the kinetic
temperature of dust (I05):
\begin{equation}\label{0}
\dot T_{d}=\frac{\sigma_{Q}^2\omega_{{\rm p}d}^2}{Q_0^2\nu_{\rm ch}}T_{d}-2\tau_{dn}^{-1}(T_d-T_n),
\end{equation}
where $\omega_{{\rm p}d}=\sqrt{4\pi Q_0^2n_d/m_d}$ is the dust plasma frequency.

Without charge fluctuations, when the collisions between dust particles conserve the energy, the equilibrium temperature of
dust is determined by interactions with the ambient gas, so that $T_d=T_n$. Random charge fluctuations provide an additional
energy source, and if the coefficient of the first (source) term in the r.h.s. of Eq. (\ref{0}) exceeds the damping rate
$2\tau_{dn}^{-1}$, then the dust temperature grows exponentially with time.

The kinetic coefficient ${\cal A}_{dd}$ (Appendix \ref{app3}) is linearly proportional to the maximum scattering angle,
which we set equal to unity. Hence, we implicitly supposed that there are sufficiently small impact parameters $\rho$ that
ensure scattering at large angles, $\chi\gtrsim1$. However, since the lower bound of impact parameters is limited by the
particle radius, $\rho\gtrsim a$, this assumption is only valid if the kinetic temperature is below a certain critical value
$T_d^{\rm cr}$. From the relation $\chi\sim Q_0^2/\rho T_d$ (Lifshitz \& Pitaevskii 1981) we readily deduce,
\begin{displaymath}
T_d^{\rm cr}\sim \frac{Q_0^2}{a}\equiv\frac{|Q_0|}{e}zT_i,
\end{displaymath}
so that the {\it actual} value of the maximum scattering angle is equal to $T_d^{\rm cr}/T_d$. Therefore, the exponential
temperature growth described by Eq. (\ref{0}) proceeds until $T_d$ reaches the critical value $T_d^{\rm cr}(\gg T_i)$. At
larger temperatures the source term in Eq. (\ref{0}) tends to a constant value (which is obtained by replacing $T_d$ with
$T_d^{\rm cr}$). The temperature growth crosses over to linear and eventually gets saturated due to the friction, which
determines the ultimate temperature of dust.

\section{Size-dependence of kinetic temperature for astrophysical grains}

Let us now determine the conditions (i) when the exponential heating described by Eq. (\ref{0}) can set in, (ii) what is the
magnitude of the ultimate kinetic temperature that can be reached by astrophysical grains due to the stochastic heating, and
(iii) what are the characteristic timescales to reach this temperature. We shall assume an isothermal ($T_n\approx
T_e\approx T_i$) quasineutral ($n_e\approx n_i$) plasma with arbitrary value of the ionization fraction $n_i/n_n$.
Furthermore, we naturally consider dust as the ``test species'' which does not change the global energy balance, and
therefore do not take into account the feedback the dust has on the ISM plasma.

(i) By substituting the above defined charge dispersion $\sigma_Q$ as well as frequencies $\omega_{{\rm p}d}$, $\nu_{\rm
ch}$, and $\tau_{dn}^{-1}$ in Eq. (\ref{0}) we obtain the heating condition,
\begin{displaymath}
\frac{|Q_0|n_d}{en_i}>\frac{16}3z(2+z)n_na^3.
\end{displaymath}
Then, by using the definition of $z$ and taking into account that in an isothermal hydrogen plasma $z\approx2.5$ (Tsytovich
1997, Fortov et al. 2005), we reduce the condition for the dust heating to the following simple formula:
\begin{equation}\label{2}
a^2\lesssim0.1\lambda_{{\rm D}}^2\frac{n_d(a)}{n_n}.
\end{equation}
This inequality shows that the heating occurs for particles below a certain critical size. The mechanism of the stochastic
acceleration is effective for collisions with particles of size $\gtrsim a$, so that the relevant dust density to be
substituted in Eq. (\ref{2}) is determined by the integral $n_d(a)=\int_a^{\infty}(dn_d/da)da$, where $dn_d/da$ is the
appropriate size distribution. Usually $dn_d/da$ decreases rather steeply and therefore the critical size should have a weak
dependence on the parameters of the ISM.

For the estimate, let us assume the MRN size distribution for dust (Mathis, Rumpl, \& Nordsieck 1977) valid for the size
range between a few dozens of {\AA} and a few tenths of $\mu$m:
\begin{equation}
\frac{dn_d}{da}=-n_{\rm H}A_{\rm MRN}a^{-3.5},\label{MRN}
\end{equation}
where $A_{\rm MRN}\sim10^{-25}$~cm$^{2.5}$ (Draine \& Lee 1984). This yields $n_d(a)/n_{\rm H}\sim0.3A_{\rm MRN}a^{-2.5}$.
Taking $m_i\sim10^{-24}$~g, $n_i\sim1$~cm$^{-3}$, $T_i\sim0.1$~eV, and assuming also $n_n\approx n_{\rm H}$, we get the
critical diameter of $\sim0.3~\mu$m, i.e., the heating can already be triggered in the submicron range. It is noteworthy
that the critical size also has a weak dependence on the stochastic properties of the charge fluctuations. For instance,
from Eq. (\ref{0}) one can readily deduce that the scaling of the critical size on the charging frequency is
$\propto\nu_{\rm ch}^{-0.2}$.

(ii) In order to determine the ultimate temperature $T_d^{\infty}$ that can be reached, we recall that when $T_d$ exceeds
the critical value $T_d^{\rm cr}$, the source term in Eq. (\ref{0}) is saturated at a constant value. Hence, we have
$T_d^{\infty}\sim 0.1(\lambda_{{\rm D}}/a)^2(n_d/n_n)T_d^{\rm cr}$. By using the definition of $z$, the critical temperature
can be rewritten as $T_d^{\rm cr}=z^2(T_i/e)^2a$. Then we finally get the following estimate for the ultimate kinetic
temperature of grains:
\begin{equation}\label{2a}
\frac{T_d^{\infty}}{T_i}\sim\frac{n_i}{n_n}\frac{\lambda_{{\rm D}}^4n_d(a)}{a}.
\end{equation}
Thus, the increase of the dust temperature relative to the temperature of gas depends very strongly on the grain size.
Assuming again the MRN distribution for $n_d(a)$ we get $T_d^{\infty}/T_i\propto a^{-3.5}$. The corresponding RMS velocity
of grains, $v_{T_d}^{\infty}=\sqrt{k_{\rm B}T_d^{\infty}/m_d}$, scales as $\propto a^{-3.25}$, whereas the dependence on the
plasma parameters is weaker, $\propto T_i^{3/2}n_i^{-1/2}$. For the parameters used above, a $30$~nm dust particle can reach
the kinetic temperature which is 5--6 orders of magnitude higher than $T_i$.

(iii) The heating timescale $\tau_{\infty}$ (i.e., when $T_d^{\infty}$ is reached) is determined by the stage of linear
temperature growth. This can be straightforwardly obtained from Eq. (\ref{0}),
\begin{displaymath}
\tau_{\infty}\sim\left(\frac{e}{Q_0}\right)^2\frac{m_d}{m_i}\frac{n_i^2\lambda_{{\rm D}}^4}{n_nv_{T_i}}.
\end{displaymath}
The heating timescale is independent of the plasma density and has rather weak dependence on other parameters,
$\tau_{\infty}\propto an_n^{-1}T_i^{-1/2}$. This estimate yields heating timescale of several Myr for $30$~nm grains.

The formalism above is rather general and is applicable to any astrophysical media, including accretion disks, circumstellar
regions and the ISM. In the next section we focus our attention on the ISM dust acceleration.

\section{Implications for ISM dust}

The ISM dust acceleration is extremely important process for shattering and coagulation of dust (see Hirashita \& Yan 2009).
We discuss both subsonic and supersonic motion of dust. In Eq. (\ref{vd_app}) (Appendix D) we rewrite Eq. (\ref{2a}) in
terms of the RMS velocity of accelerated grains assuming the MRN dust size distribution, Eq. (\ref{MRN}),
\begin{equation}
v_{T_d}^{\infty}\approx\frac{6\times 10^{2}}{(\alpha n_i)^{1/2}}\left(\frac{T_{i}}{100{~\rm K}}\right)^{3/2}
\left(\frac{a}{10^{-6}{~\rm cm}}\right)^{-3.25}{\rm cm/ s},\label{vd}
\end{equation}
where where $n_i$ is in cm$^{-3}$ and the renormalizing factor $\alpha$ for the damping rate in the subsonic and supersonic
regimes is given by Eq. (\ref{alpha1}) and (\ref{alpha2}), respectively. Note that in the latter case $\alpha$ is the
function of dust velocity and hence Eq. (\ref{vd}) should be resolved for $v_{T_d}^{\infty}$.

\begin{figure}
\includegraphics[width=0.5\textwidth]{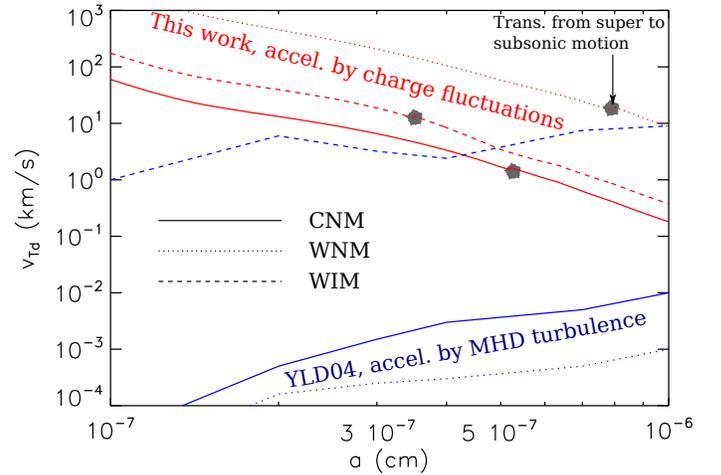}
\caption{RMS velocity of dust grain for the parameters of ISM phases given in Table 1. The grain size distribution function
from Weingartner \& Draine (2001) is adopted. Diamond symbols denote the transition from supersonic to subsonic motion of dust.}
\label{f2}
\end{figure}

In Figure~\ref{f2} we compare the mechanism presented in the paper with the calculations of the velocities of interstellar
grains obtained in YLD04. We see that for sufficiently large grains the acceleration due to MHD fluctuations is stronger. At
the same time, the new mechanism of acceleration associated with charge fluctuations is more efficient for
$a\lesssim10^{-5}$~cm.

The efficiency of the charge-fluctuation-induced acceleration increases for smaller grains. However, the limitation of our
treatment above is that the grain charging was considered as a continuous process [see Eq. (\ref{24})]. Formally, this
approach is not applicable to the situation when the mean grain charge $Q_0$ is so small that each act of losing or
acquiring an elementary charge results in substantial discrete fluctuations. However, we believe that even in this case we
may get right order of magnitude estimate of grain acceleration. Further research with Monte-Carlo simulations instead of
using Fokker-Plank approach should test the latter conjecture.

Comparing the Figures~\ref{f1} and \ref{f2} we see the limited nature of our quantitative calculations. For grains larger
than $\sim3\times 10^{-7}$~cm the interactions decrease due to the less efficient momentum deposition as grains mostly feel
the electric field of grains smaller than their size. On the other hand, the charge fluctuations become large for grains
appreciably smaller than this size and this makes the treatment of the problem with the Fokker-Planck equation not precise.
Nevertheless, even taking into account the pilot nature of our estimates it is clear that the process of acceleration is
potentially very important for small grains.

\section{Discussion}

We showed that there exists a novel powerful mechanism of charge-fluctuation-induced acceleration operating in astrophysical
plasmas. This mechanism, which occurs to be more efficient for smaller grains, is based on the charge fluctuations that
grains experience during the mutual Coulomb collisions. As the result, the kinetic energy of grains is not conserved in the
collisions, which causes the stochastic heating analogous to the second-order Fermi acceleration.

This acceleration mechanism is a generic plasma process which can operate in very different environments ranging from the
ISM to laboratory gas discharges. It is a remarkable manifestation of the intrinsic non-equilibrium (energetic openness) of
dusty plasmas (Tsytovich 1997, Fortov 2005): Even if electrons and ions themselves were in detailed equilibrium, their
absorption and the subsequent recombination on dust grains cannot be balanced by the corresponding inverse process (of the
ionization and emission from grains) due to its apparent inefficiency -- temperatures necessary for that should be
enormously high. Therefore, the process of dust charging goes only in one direction, resulting in the imbalanced plasma flux
on the grain surface. The dust dynamics gets coupled to the charging process due to charge fluctuations (which are
inevitable due to the discreteness of the elementary charge). Thus, in terms of the energy balance, the mechanism of
charge-fluctuation-induced acceleration is based on the conversion of the energy flux (associated with the plasma flux on
grains) into the kinetic energy of dust.



One should point out that for typical ISM conditions the situation is very different from the equilibrium one: The plasma
temperature is much larger that the dust material temperature ($T_{d{\rm m}}$) due to efficient radiative cooling of the
dust. Moreover, the ions and electrons are usually produced by the ISM UV and/or cosmic ray ionization. Grains themselves
emit photoelectrons interacting with UV photons. All these non-equilibrium effects can alter charge fluctuations and
therefore additionally contribute to our acceleration mechanism.

In particular environments, e.g., within quiescent dark clouds and protoplanetary disks with suppressed turbulence, the
mechanism we discussed in the paper may be dominant for grains substantially larger than $10^{-5}$~cm. However, we notice
the rapid decrease of the mechanism efficiency with the grain size.

The treatment presented in this paper is limited in two respects. First, the Fokker-Planck approach presumes relatively
small energy variation (\ref{25}) occurring in each Coulomb collision due to charge fluctuations. If this variation becomes
large, such approach (describing the evolution of the velocity distribution function) is no longer applicable. On the other
hand, the mean energy evolution -- which is essential for this paper -- is still described by Eq. (\ref{0}). Second, the
direct mechanical encounters of grains cannot be treated within the formalism. When such collisions start dominating (at
large $T_d$) the additional dissipation associated with grain deformation, shattering, etc. should be taken into account.

In future, we plan to extend the mechanism for the PAH population of grains, explicitly taking into account the substantial
variation of grain charge in the process of individual collision with an electron or ion (remind that in this paper we
assumed small charge fluctuations, which allowed us to obtain analytically tractable results). The fact that the efficiency
of grain acceleration increases with the decrease of grain size makes the process very efficient. Extrapolating our results
to PAH particles we may expect their collisions to be very frequent. We note that the additional motivation for the PAH
studies comes from the fact that they are responsible for a component of CMB foreground radiation (Draine \& Lazarian 1998,
see also Hoang et al. 2010).

Furthermore, we plan to analyze the impact of the charge-fluctuation-induced acceleration on the dust coagulation processes
occurring in the ISM and other astrophysical environments, e.g. protostellar accretion disks. It is already clear that the
new mechanism can induce coagulation and shattering which have not been considered in the literature before. Together with
the MHD turbulence acceleration, our mechanism testifies that the astrophysical grains achieve velocities much larger than
those arising from the Brownian motion.

\section{Summary}

Our major results can be summarized as follows:\\
1. The change of the grain charge in the process of the grain-grain Coulomb collisions results in second-order Fermi
acceleration.\\
2. The acceleration is most efficient for small, i.e., less than $\sim10^{-6}$~cm grains.\\
3. The process of acceleration should be accounted for correct description of the evolution of the smallest grains.

\acknowledgments
AL and TH acknowledge the support of the NSF-funded Center for Magnetic Self-Organizatiaon (CMSO) and the NSF grant AST 0507164.

\appendix

\section{A. Collision operator for dust-gas interactions}
\label{app1}

The interaction of individual dust particles with the background neutral gas is described by the Langevin equation (see Van
Kampen 1981),
\begin{displaymath}
\dot{\bf p}=-\tau_{dn}^{-1}{\bf p}+{\bf L}(t),
\end{displaymath}
where $\tau_{dn}^{-1}$ is given by Eq. (\ref{tau_fr}). The random Langevin force has zero average, $\langle{\bf
L}(t)\rangle=0$, and is properly normalized, $\langle{\bf L}(t){\bf L}(t+\tau)\rangle=2\tau_{dn}^{-1}m_dk_{\rm
B}T_n\delta(\tau)$, to satisfy the fluctuation-dissipation theorem.

The formalism of the Langevin equation is equivalent to the Fokker-Planck approach which describes evolution of the velocity
distribution function (Lifshitz \& Pitaevskii 1981, Van Kampen 1981). The latter approach is solely determined by the first
and second Fokker-Plank coefficients, ${\cal A}=\langle\delta{\bf p}\rangle/\delta t$ and ${\cal B}=\langle(\delta{\bf
p})^2\rangle/2\delta t$ (where $\langle\ldots\rangle$ denotes the averaging over many collisions occurring over the time
period $\delta t$). They play the role of the mobility and diffusion coefficients in the velocity space, respectively, and
for processes that allow stable equilibrium (e.g., for dust-neutral collisions) they are interrelated. The resulting
collision operator for dust-neutral collisions has the following differential form (Lifshitz \& Pitaevskii 1981, Van Kampen
1981),
\begin{displaymath}
{\rm St}_{dn}f_d({\bf p})={\cal B}_{dn}\frac{\partial}{\partial{\bf p}}\left(\frac{{\bf p}}{mT_n}f_d+ \frac{\partial
f_d}{\partial{\bf p}}\right),
\end{displaymath}
where ${\cal B}_{dn}=\tau_{dn}^{-1}m_dk_{\rm B}T_n$ is the second Fokker-Plank coefficient for dust-neutral collisions. The
integration immediately yields
\begin{equation}\label{int1}
\int\frac{p^2}{m_d}{\rm St}_{dn}f_{\rm M}d{\bf p}=-6\tau_{dn}^{-1}k_{\rm B}\left(T_d-T_n\right).
\end{equation}

\section{B. Collision operator for dust-dust interactions}
\label{app2}

We employ the most general form of the collision operator for binary collisions,
\begin{equation}
{\rm St}_d f({\bf p})=\int\Big[w({\bf p}',{\bf p}_1';\:{\bf p},{\bf p}_1)f({\bf p}')f({\bf p}_1')-
w({\bf p},{\bf p}_1;\:{\bf p}',{\bf p}_1')f({\bf p})f({\bf p}_1)\Big]d{\bf p}_1d{\bf p}'d{\bf p}_1'.\label{St_n}
\end{equation}
This form allows for the energy non-conservation in the mutual dust collisions and therefore does not require the unitarity
relation to be satisfied [which reduces Eq. (\ref{St_n}) to the canonical Boltzmann form (Lifshitz \& Pitaevskii 1981)].
Here, $w({\bf p},{\bf p}_1;\:{\bf p}',{\bf p}_1')$ is a probability function for a pair of colliding particles with momenta
${\bf p}$ and ${\bf p}_1$ to acquire momenta ${\bf p}'$ and ${\bf p}_1'$, respectively, after the collision. Equation
(\ref{St_n}) counts for all possible transitions $({\bf p}',{\bf p}_1')\to({\bf p},{\bf p}_1)$ and then is averaged over
${\bf p}_1$. Function $w$ can be determined by solving a mechanical problem of the binary scattering with given interaction
between the particles.

Assuming the Maxwellian distribution, $f({\bf p})=f_{\rm M}({\bf p})$, it is convenient to introduce the ``binary''
distribution function of dust, $F({\bf p}_{\rm c},{\bf p}_{\rm r})\equiv f_{\rm M}({\bf p}_1)f_{\rm M}({\bf p}_2)$. We also
define the probability function $W({\bf p}_{\rm c},{\bf p}_{\rm r};\:{\bf q})\equiv w({\bf p},{\bf p}_1;\:{\bf p}',{\bf
p}_1')$ for the relative momentum exchange ${\bf p}_{\rm r}\to{\bf p}_{\rm r}+{\bf q}$ and after the integration of Eq.
(\ref{St_n}) obtain (I05),
\begin{equation}
\int\frac{p^2}{m_d}{\rm St}_{dd} f_{\rm M}d{\bf p}=
\frac14\int\frac{p_{\rm r}^2}{m_d}\Big[W({\bf p}_{\rm c},{\bf p}_{\rm r}-{\bf q};\:{\bf q})F({\bf p}_{\rm c},{\bf p}_{\rm r}-{\bf q})-
W({\bf p}_{\rm c},{\bf p}_{\rm r};\:{\bf q})F({\bf p}_{\rm c},{\bf p}_{\rm r})\Big]d{\bf p}_{\rm c}d{\bf p}_{\rm r}d{\bf q}.\label{7}
\end{equation}

In a homogeneous and isotropic medium, the collision probability depends only on the absolute value of the relative
momentum, $p_{\rm r}$, and does not depend on ${\bf p}_{\rm c}$. Thus, $W=W(p_{\rm r};\:\delta q)$, where we introduced the
absolute value for the inelastic momentum exchange, $\delta q=|{\bf p}_{\rm r}+{\bf q}|-p_{\rm r}$. For weakly inelastic
collisions with $\delta q\ll p_{\rm r}$ we have
\begin{equation}\label{20a}
\delta q\approx2\frac{m_d\delta\varepsilon_{\rm r}}{p_{\rm r}}.
\end{equation}
Furthermore, the integrand in Eq. (\ref{7}) can be expanded into a series over $\delta q$. Retaining the linear and
quadratic terms and integrating in parts, we obtain
\begin{equation}\label{8}
\int\frac{p^2}{m_d}{\rm St}_{dd}f_{\rm M}d{\bf p}\approx\frac1{2m_d}\int (p_{\rm r}{\cal A}_{dd}+
{\cal B}_{dd})f_{\rm M}d{\bf p}_{\rm r},
\end{equation}
where ${\cal A}_{dd}(p_{\rm r})=\int\delta qWd\delta q$ and ${\cal B}_{dd}(p_{\rm r})=\frac12\int(\delta q)^2Wd\delta q$ are
analogous to the Fokker-Planck coefficients for {\it inelastic} dust-dust collisions (I05). The coefficients can be
expressed in terms of the binary collision cross section $\Sigma$ in the following form:
\begin{equation}\label{FP_coeff}
\begin{array}{l}
{\cal A}_{dd}=\int\delta q\; n_dv_{\rm r}d\Sigma,\\[.4cm]
{\cal B}_{dd}=\frac12\int(\delta q)^2n_dv_{\rm r}d\Sigma.
\end{array}
\end{equation}
Using Eqs (\ref{25}) and (\ref{20a}) we obtain that $\delta q=O(\sigma_Q^2)$. Therefore, ${\cal A}_{dd}=O(\sigma_Q^2)$ and
${\cal B}_{dd}=O(\sigma_Q^4)$, i.e., for weakly fluctuating charges the mean energy variation is determined by coefficient
${\cal A}_{dd}$ only.

\section{C. Calculation of ${\cal A}_{dd}$}
\label{app3}

In order to calculate the kinetic coefficient ${\cal A}_{dd}$, we assume the Coulomb interaction between charged grains (the
main contribution to ${\cal A}_{dd}$ comes from the interaction at shorter distances, see I05). By substituting
$d\varphi_0/dr=-Q_0/r^2$ in Eq. (\ref{25}) and integrating along the trajectory $r(t)=\sqrt{\rho^2+(v_{\rm r}t)^2}$, we
obtain the energy variation $\delta\varepsilon_{\rm r}\sim(\sigma_Q^2Q_0^2/\rho^3\nu_{\rm ch}p_{\rm r})$ (here $\rho$ is the
impact parameter of colliding particles, for simplicity we consider the scattering at small angles) and the corresponding
momentum variation $\delta q$ [using Eq. (\ref{20a})]. Finally, we substitute the result in the first equation
(\ref{FP_coeff}), where we use the relation between the scattering angle and the impact parameter in the Coulomb limit,
$\chi\sim Q_0^2/\rho\varepsilon_{\rm r}$, and employ the Coulomb differential cross section,
$d\Sigma\sim(Q_0^4/\varepsilon_{\rm r}^2 \chi^3)d\chi$ (Landau \& Lifshitz 1976). By integrating over $0\leq\chi\leq1$ we
derive,
\begin{equation}\label{27}
{\cal A}_{dd}\sim\frac{\sigma_Q^2\omega_{{\rm p}d}^2p_{\rm r}}{Q_0^2\nu_{\rm ch}}.
\end{equation}

\section{D. Numerical estimate for RMS grain velocity}\label{app4}

The ultimate kinetic temperature of grains, $T_d^{\infty}$, is given by Eq. (\ref{2a}). It is determined by the plasma Debye
length, \bea \lambda_{{\rm D}}=\left(\frac{k_{B}T_{i}}{4\pi e^{2}n_{i}}\right)^{1/2}\approx\frac{63}{n_i^{1/2}}
\left(\frac{T_{i}}{100 {~\rm K}}\right)^{1/2}{\rm cm},\label{eq2} \ena where $n_i$ is in cm$^{-3}$. Let us assume the MRN
size distribution for dust, Eq. (\ref{MRN}). Taking into account that the relevant dust density is
$n_d(a)=\int_a^{\infty}(dn_d/da)da$, assuming that $n_{n}\sim n_{\rm H}$, and using Eq. (\ref{eq2}) we get \bea
\frac{T_d^{\infty}}{T_i}\approx \left(\frac{4\times 10^{9}}{\alpha n_{i}}\right)\left(\frac{T_{i}}{100 {~\rm
K}}\right)^{2}\left(\frac{a}{10^{-8}{~\rm cm}}\right)^{-3.5},\label{td} \ena where the renormalizing factor $\alpha$ for the
subsonic and supersonic regimes is given by Eq. (\ref{alpha1}) and (\ref{alpha2}), respectively. The corresponding mean
squared velocity of grains is \bea (v_{T_d}^{\infty})^{2}={\frac{k_{\rm B}T_d^{\infty}}{m_d}}={T_{d}^{\infty}}\times
10^{7}\left(\frac{a}{10^{-8}{~\rm cm}}\right)^{-3}({\rm cm/s})^2,\label{vd2} \ena where the dust mass $m_{d}=(4\pi/3)\rho
a^{3}\approx 10^{-23} (a/10^{-8}{~\rm cm})^{3}$ is used. By substituting (\ref{td}) into (\ref{vd2}) we finally get \bea
v_{T_d}^{\infty}\approx\frac{6\times 10^{2}}{(\alpha n_i)^{1/2}}\left(\frac{T_{i}}{100{~\rm K}}\right)^{3/2}
\left(\frac{a}{10^{-6}{~\rm cm}}\right)^{-3.25}{\rm cm/ s}.\label{vd_app} \ena


\begin{thebibliography}{}

\bibitem[Allamandola et al.(1989)]{1989ApJS...71..733A} Allamandola, L.~J., Tielens, A.~G.~G.~M., \& Barker, J.~R. 1989,
    \apjs, 71, 733
\bibitem{deAngelis05} de Angelis, U., Ivlev, A. V., Morfill, G. E., \& Tsytovich, V. N. 2005, Phys. Plasmas, 12,
    052301

\bibitem[Cho \& Lazarian(2003)]{2003MNRAS.345..325C} Cho, J., \& Lazarian, A. 2003, \mnras, 345, 325

\bibitem[Cho \& Lazarian(2002)]{2002PhRvL..88x5001C} Cho, J., \& Lazarian, A. 2002, Physical Review Letters, 88, 245001


\bibitem[Draine(2009)]{2009EAS....35..245D} Draine, B. T. 2009, EAS Publications Series, 35, 245
\bibitem[]{Draine+Li_1984} Draine, B. T., \& Lee, H. M. 1984, \apj, 285, 89

\bibitem[]{Draine+Li_2007} Draine, B. T., \& Li, A. 2007, \apj, 657, 810

\bibitem[Draine(1985)]{1985prpl.conf..621D} Draine, B. T. 1985, Protostars and Planets II, 621

\bibitem[Draine(2003)]{2003ARA&A..41..241D} Draine, B. T. 2003, \araa, 41, 241
\bibitem[]{DrSu87} Draine, B. T., \& Sutin, B. 1987, ApJ, 320, 803
\bibitem[Draine \& Salpeter(1979)]{1979ApJ...231..438D} Draine, B. T., \& Salpeter, E. E. 1979, \apj, 231, 438

\bibitem{Fortov05} Fortov, V. E., Ivlev, A. V., Khrapak, S. A., Khrapak, A. G., \& Morfill, G. E. 2005, Phys. Rep.,
    421, 1

\bibitem[Fraisse et al.(2009)]{2009AIPC.1141..265F} Fraisse, A. A., et al. 2009, American Institute of Physics Conference
    Series, 1141, 265

\bibitem{Goree94} Goree, J. 1994, Plasma Sources Sci. Technol., 3, 400

\bibitem[]{Hoang10} Hoang, T., Draine, B.~T., \& Lazarian, A. 2010, \apj, 715, 1462

\bibitem{Ivlev04} Ivlev, A. V., Zhdanov, S. K., Klumov, B. A., Tsytovich, V. N., de Angelis, U., Morfill, G. E. 2004,
    Phys. Rev. E, 70, 066401

\bibitem{Ivlev05} Ivlev, A. V., Zhdanov, S. K., Klumov, B. A., \& Morfill, G. E. 2005, Phys. Plasmas 12, 092104

\bibitem{Khrapak99} Khrapak, S. A., Nefedov, A. P., Petrov, O. F., \& Vaulina, O. S. 1999, Phys. Rev. E 59, 6017

\bibitem{Landau_Mechanics} Landau, L. D., \& Lifshitz, E. M. {\it Mechanics} (Pergamon, Oxford, 1976)

\bibitem[Lazarian \& Finkbeiner(2003)]{2003NewAR..47.1107L} Lazarian, A., \& Finkbeiner, D. 2003, New Astronomy Review, 47,
    1107
\bibitem[Lazarian \& Yan(2002)]{2002ApJ...566L.105L} Lazarian, A., \& Yan, H. 2002, \apjl, 566, L105
\bibitem[Leger \& Puget(1984)]{1984A&A...137L...5L} Leger, A., \& Puget, J. L. 1984, \aap, 137, L5
\bibitem{Landau_Kinetics} Lifshitz, E. M., \& Pitaevskii, L. P. {\it Physical Kinetics} (Pergamon, Oxford, 1981)
\bibitem[]{Mathis, Rumpl, Nordsieck 1977} Mathis, J. S.,  Rumpl, W., \& Nordsieck, K. H. 1977, \apj, 217, 425
\bibitem{Matsoukas97} Matsoukas, T., \& Russell, M. 1997, Phys. Rev. E, 55, 991
\bibitem[Purcell(1969)]{1969Phy....41..100P} Purcell, E. M. 1969, Physica, 41, 100

\bibitem[Tielens(2005)]{2005pcim.book.....T} Tielens, A.~G.~G.~M. 2005, The Physics and Chemistry of the Interstellar
    Medium, by A.~G.~G.~M.~Tielens, pp.~.~ISBN 0521826349.~Cambridge, UK

\bibitem{Tsytovich97} Tsytovich, V. N. Phys. Usp. 1997, 40, 53

\bibitem{Tsytovich08} Tsytovich, V. N., Morfill, G. E., Vladimirov, S. V., \&  Thomas, H. M. {\it Elementary physics of
    complex plasmas} (Springer, Berlin, 2008)

\bibitem{Spitzer} Spitzer, L. {\it Physical Processes in the interstellar medium} (New York, 1978)
\bibitem{VanKampen} van Kampen, N. G. {\it Stochastic Processes in Physics and Chemistry} (Elsevier, Amsterdam, 1981)

\bibitem{Yan04} Yan, H., Lazarian, A., \& Draine, B. T. 2004, ApJ, 616, 895
\bibitem[Yan \& Lazarian(2003)]{2003ApJ...592L..33Y} Yan, H., \& Lazarian, A. 2003, \apjl, 592, L33

\bibitem[Weingartner \& Draine(2001)]{2001ApJ...563..842W} Weingartner, J. C., \& Draine, B. T. 2001, \apj, 563, 842

\bibitem{WhittetBook} D. C. B. Whittet, {\it Dust in the Galactic Environment} (IOP Publ., Bristol, 2003).


\end{thebibliography}
\end{document}